# PIP-II Injector Test's Low Energy Beam Transport: Commissioning and Selected Measurements

A. Shemyakin[1, a)], M. Alvarez[1], R. Andrews[1], J.-P. Carneiro[1], A. Chen[1], R. D'Arcy[2], B. Hanna[1], L. Prost[1], V. Scarpine[1], C. Wiesner[3]

[1]*Fermilab, Batavia, IL 60510, USA*
[2] *University College London, London, WC1E 6BT, UK*
[3]*IAP, Goethe University, Frankfurt am Main, Germany*

a)Corresponding author: shemyakin@fnal.gov

**Abstract.** The PIP2IT test accelerator is under construction at Fermilab. Its ion source and Low Energy Beam Transport (LEBT) in its initial (straight) configuration have been commissioned to full specification parameters. This paper introduces the LEBT design and summarizes the outcome of the commissioning activities.

## INTRODUCTION

The Proton Improvement Plan II (PIP-II) is a program of upgrades to the Fermilab injection complex [1]. As presently envisioned, the PIP-II core is a new 2 mA, 800 MeV H$^-$ CW superconducting linac working initially in a pulse mode (0.55 ms, 20 Hz). To validate the concept of such machine's front-end, a test accelerator (a.k.a. PIP-II Injector Test, or PIP2IT; previously referred to as PXIE) is under construction [2]. It includes a 15 mA DC, 30 keV H$^-$ ion source, a 2 m-long LEBT, a 2.1 MeV CW RFQ, a Medium Energy Beam Transport (MEBT), 2 cryomodules increasing the beam energy to ~25 MeV, and a High Energy Beam Transport section that takes the beam to a dump.

The LEBT was studied in several configurations, differing primarily by the placement of diagnostics. In its present, straight configuration, the LEBT has been fully commissioned and successfully used for injection into the RFQ. The paper describes the LEBT design, highlights several commissioning aspects, and presents two specific measurements: neutralizing ions production and intensity of the fast particles produced through H$^-$ electron detachment.

## LEBT DESIGN AND LAYOUT

### Specifications and Scheme

The primary requirements for the LEBT are summarized in TABLE 1. Its nominal mode of operation is DC, but for commissioning purposes, the LEBT is required to provide a wide range of duty factors.

TABLE 1: LEBT beam parameters and main requirements.

| Parameter | Value | Unit |
|---|---|---|
| Kinetic energy | 30 | keV |
| Nominal/Maximum beam current, DC | 5/10 | mA |
| Output transverse emittance over 2-5 mA current range | < 0.18 | µm |
| Typical pressure | ~$10^{-6}$ | Torr |
| Chopping frequency, max | 60 | Hz |
| Pulse length | 1-16600 | µs |

The choice of the LEBT scheme is discussed in detail in Ref. [3]. Its peculiarity is in the neutralization pattern: nearly full neutralization through the upstream portion of the beam line, followed by a non-neutralized region over the last ~1m from the chopper to the RFQ (Fig. 1a). The main reason for this choice is to maintain as good a vacuum as possible in the RFQ, while avoiding long transient times due to space charge neutralization when chopping [4]. Note that such a scheme is possible because the beam perveance is sufficiently small.

In addition, for PIP-II, it is envisioned to install two ion sources (as shown schematically in Fig. 1b) in order to maximize the beam availability. For flexibility in the beam optics solutions providing the transport through the switching dipole, the LEBT is composed of 3 solenoids (for each leg). While there is no plan of installing a second source at PIP2IT, the switching dipole magnet will be installed and commissioned in 2016. Meanwhile, RFQ commissioning is being conducted in a straight LEBT configuration, and this paper describes results for the latter.

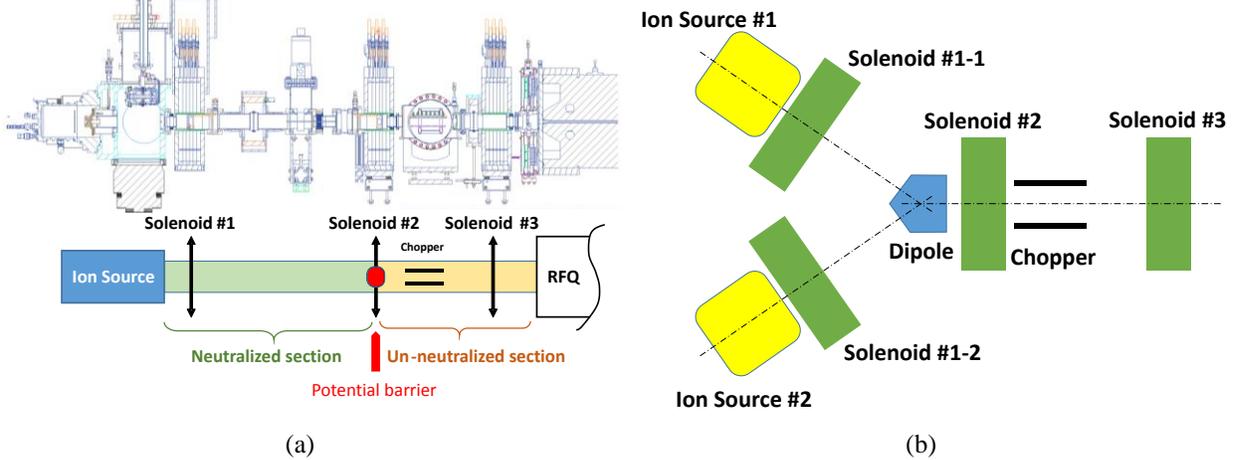

**FIGURE 1.** Concept of the LEBT. (a) Neutralization scheme. (b) Schematic of the PIP-II LEBT with two ion sources.

## Beam Line Components

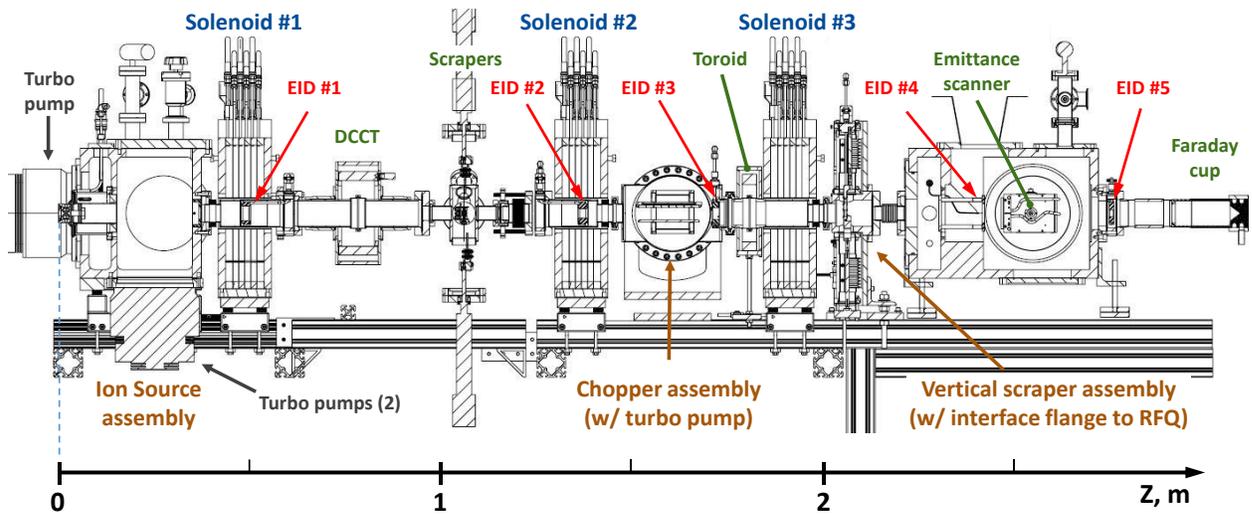

FIGURE shows the model of the ion source and LEBT in its straight configuration. The ion source is a non-cesiated, TRIUMF-type (filament driven) 15-mA, 30 keV H$^-$ volume-cusp ion source from D-Pace, Inc. [5]. While the source is designed to be operated DC, a modulation circuit is added to its extraction electrode, which allows pulses from 5 μs to 16 ms to be generated, at a frequency of up to 60 Hz with rise and fall times of the order of 1 μs. The source assembly includes 2 dipole correctors at the exit of the ion source ground electrode.

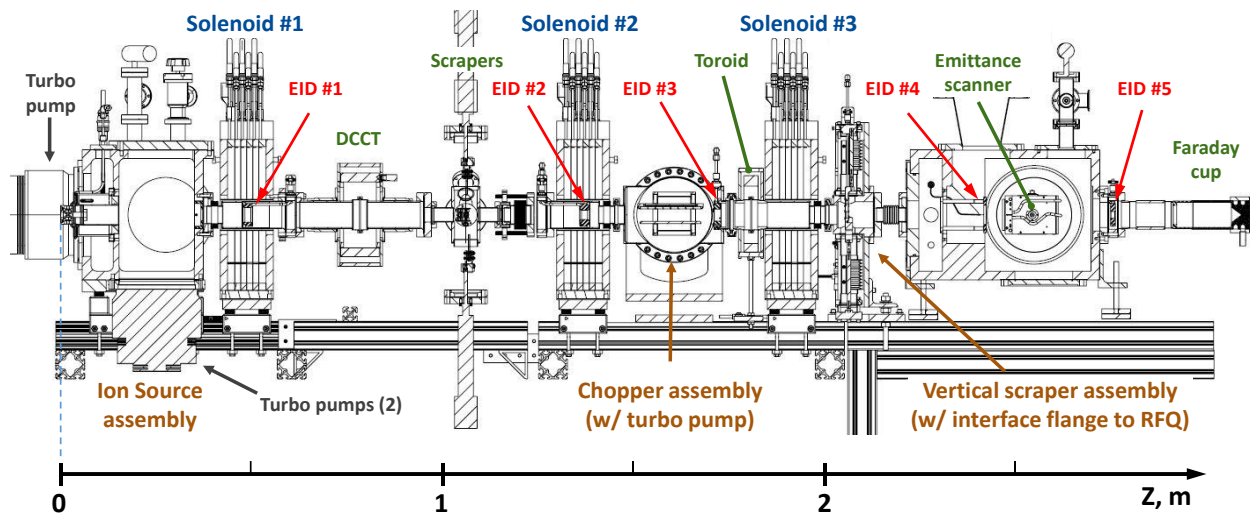

**FIGURE 2.** PIP2IT ion source and LEBT CAD model (vertical cross section).

The ground electrode effectively divides the vacuum enclosure into two chambers with limited conductance for gases. According to simulations performed with the MOLFLOW code [6], the pressure in the extraction region is ~10 mTorr and < 0.1 mTorr in the downstream chamber for a $H_2$ gas flow of 15 sccm.

The ion source assembly is followed by (in that order): solenoid #1, a DCCT, a beam stop (for personnel protection), a vacuum valve, solenoid #2, a chopper, a current transformer (a.k.a. toroid), solenoid #3, and a 'LEBT scraper' assembly. During the LEBT studies, instead of the RFQ, the beam propagates into a diagnostics vacuum chamber and a deep Faraday cup. An emittance scanner is installed either on the diagnostics vacuum chamber or on the ion source vacuum box. The diagnostics box can be rotated 90° in order to measure the phase space of the beam in both orthogonal directions.

The solenoids produce a peak field of 0.62 T at 300 A max current and a maximum field integral on axis of 0.034 $T^2$·m. Each solenoid includes a pair of steering dipole coils within its bore. At the peak current of 15 A, the integrated deflecting field is 0.5 mT·m. There are water-cooled electrically isolated diaphragms (EID) in the $1^{st}$ and $2^{nd}$ solenoid, which can be biased to up to ±100 V.

The chopper assembly is an electrostatic kicker followed by another EID (#3). The kicker plates, 32 mm apart, are 16 cm long and 5 cm wide. The bottom ("kicking") plate, made of stainless steel, is high-voltage isolated. The beam is inhibited when this plate sits at -5 kV and goes through when it is brought towards zero (or -300 V). The pulse width can be varied from 1 µs to 16 ms, at a rate of up to 60 Hz, with rise and fall times of < 100 ns, independently from the ion source modulator. The top plate of the kicker also acts as the absorber. The surface exposed to the beam is made of a molybdenum alloy TZM. It is divided in 8 strips along the beam propagation for stress relief. The TZM strips are bolted onto a water-cooled copper frame, which is electrically isolated, thus allowing currents to be read back. Compared to many choppers at other facilities, which are designed to be as close as possible to the RFQ and, as a result, compact, the PIP2IT chopper is on purpose relatively bulky to be robust.

In addition to providing a pulsed beam, the LEBT chopper is the primary beam–inhibiting device for the Machine Protection System (MPS) of PIP2IT. To improve its reliability, the kicking plate voltage is read through a separate connector and compared with the request. The ion source bias voltage is crowbarred in a case of discrepancy.

Just downstream of solenoid #3, there is a movable electrode called "LEBT scraper". It is a water-cooled and electrically isolated copper paddle attached to a stepper motor drive (vertical direction). There are two round apertures, with TZM inserts on the face oriented toward the beam, and a larger "D-shaped" hole. The smallest (3.5 mm diam.) round aperture provides a 'pencil-like' beam downstream. The larger (9 mm) round aperture is intended to protect the RFQ vanes during normal operation and scrape the beam tails. The "D-shaped" hole allows for passing the entire beam through. Its horizontal edge is also used to measure the beam profile.

Vacuum is obtained with four 1000-l/s turbo pumps. Three of them are located on the ion source vacuum chamber (one on the upstream chamber where extraction takes place and two on the downstream chamber) and another one is on the chopper assembly. Downstream of the ion source chamber, the pressure gradually drops to ≤ $10^{-6}$ Torr. Vacuum pressure in the chopper assembly remains in the low $10^{-7}$ Torr with 10 mA DC beam being sent to the absorber plate. According to RGA measurements, this pressure is dominated by recombination of the beam ions into hydrogen at the absorber surface. Comparison with vacuum calculations indicates that ~50% of the incoming ions recombine with the remaining part probably diffusing into the bulk.

## Instrumentation

The LEBT instruments can be divided according to their functionalities into beam current, beam position and transverse profile, and transverse emittance diagnostics. Most of them remain in the beam line, although some were installed and used for LEBT commissioning only.

For total beam current measurements, the LEBT utilizes a Bergoz DC Current Transformer (DCCT), located after solenoid #1, a Pearson 7655 beam current toroid located directly after the chopper, and a water-cooled deep Faraday cup at the end of the beam line. The DCCT provides a reliable current measurement for pulses longer than 0.8 ms (at the end of the pulse). For beam current measurements of transverse tails, already mentioned diaphragms (EID) are used. One of them (EID #4) is located just in front of the Faraday cup. The current read back by the EIDs is a true representation of the beam lost onto them when biased positively (> +40V). The EIDs as well as the LEBT scraper apertures are also used in conjunction with the solenoid dipole correctors to align the beam and in some instances to measure the beam size [7].

Integrated beam current density profiles near the end of the LEBT can be obtained by steering the beam across the edge of the D-shaped hole or, conversely, by moving the LEBT scraper (vertical direction only) while keeping the beam fixed, and recording the intercepted and transmitted currents. The beam size can be obtained either by fitting the data to analytic formulae for a specific current density distribution (i.e. Gaussian or uniform) or by a direct rms calculation.

In one of the commissioning configuration of the LEBT, a set of 4 scrapers (2H + 2V) was installed as shown in Fig.2 in the location normally occupied by a beam stop. Each scraper was an independently moving, electrically isolated TZM plate cooled by radiation. The scrapers were used for measuring the beam profiles in that location similarly to the LEBT scraper measurement. An example of a scraper scan and associated profile is shown in Fig.3.

A water-cooled Allison-type emittance scanner has been designed for the PIP2IT LEBT. The device is a modified version from previous LBNL/SNS designs with the addition of water-cooled front slits (and other minor differences). A detailed description of the scanner can be found in reference [8] along with its calibration procedure and error analysis. The emittance measurement uncertainty is found to be <5%.

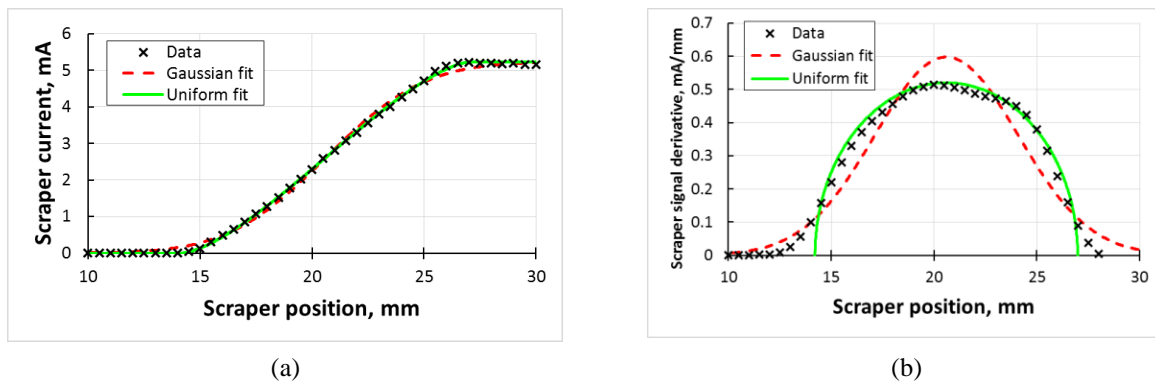

(a) (b)

**FIGURE 3**. Horizontal beam size measurements with a scraper. (a) - current measured from the scraper plate as a function of the scraper position; (b) – density projection on the horizontal axis reconstructed as a derivative of the data in (a) over position. The data are recorded 1 ms from the front of a 1.5 ms modulator pulse.

## Beam optics simulations

The beam dynamics simulations during the design and commissioning of the LEBT employed in most cases the code TRACK [9]. Typically, simulations assume that the beam is fully neutralized up to EID #2 and completely un-neutralized afterward. Generation of the initial particle distribution is based on phase space measurements at the exit of the ion source. An example of the beam rms size simulation is shown in Fig. 4.

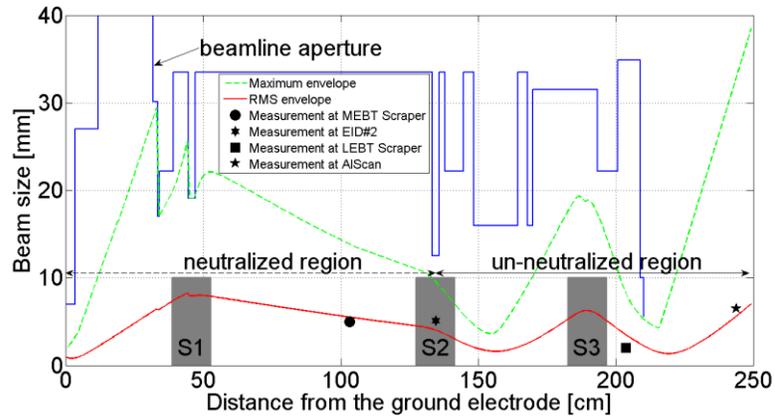

**FIGURE 4**. Comparison of the rms beam size measured with different instruments along the beam line for the same beam settings (symbols) with simulations (red line). The blue contour represents the vacuum chamber aperture. The beam current out of the LEBT is 5 mA. In the longitudinal coordinate, z=0 corresponds to the location of the ground electrode.

A detailed discussion of simulations and comparison with measurements will be reported in Ref. [10].

## COMMISSIONING

The main goals of commissioning were to understand the optics of the line, optimize beam parameters for injection into the RFQ, and prepare hardware and procedures for operation of future sections downstream.

Most of measurements were made in a pulsed mode to minimize chances for equipment damage during tuning. Typically, the modulator formed several – ms pulses, and the chopper was cutting out the initial 0.7 – 1 ms of the pulse, where beam neutralization upstream of Solenoid #2 is taking place. Measurements in high-duty and DC modes were performed mainly to verify thermal capabilities of elements and study the beam stability.

Calibrations of solenoids and dipole correctors were verified through beam measurements using EIDs, scrapers, and the emittance scanner. The phase portraits were recorded for a large variety of ion source settings.

The main result of the LEBT commissioning was the successful demonstration of all its required capabilities: stable operation in both pulsed and DC modes with currents up to 10 mA; a beam emittance at the LEBT exit of 0.13 µm (rms, norm.) at the nominal current of 5 mA and the Twiss functions at the future location of the RFQ vanes being close to targets; and good vacuum at the LEBT exit ($< 3 \cdot 10^{-7}$ Torr) in all beam modes of operation. In addition, the measurements indicated clearly that the scheme presented in Fig.1a works [11].

Lately, the 5 mA beam formed by the LEBT was accelerated in the RFQ with transmission of >95%. Details of the RFQ commissioning are outside the scope of this paper and are discussed elsewhere [12]. Below, we highlight several commissioning results of the LEBT in its straight configuration.

### Transmission through the first solenoid

After installing the first solenoid, it was realized that the beam exiting the ion source was more divergent than expected, resulting in losses at the exit of the ion source box rather than at the EID #1. Therefore, the current measured on EID #1 does not account for the entire beam loss upstream of the DCCT.

In the LEBT final configuration, the emittance scanner was moved to the ion source vacuum chamber (hence before this aperture limitation). In this location, one can use the image integral from the phase space portraits (Fig.5a) as an indirect measurement of the beam current coming out of the ion source. The coefficient of proportionality between the phase space integral and the beam current was measured when the scanner was at the end of the LEBT by comparison with the Faraday Cup current for the same conditions. Assuming that this calibration did not change when moving the Allison scanner into the ion source box, the beam current as a function of the ion source extraction electrode voltage is plotted in Fig. 5b.

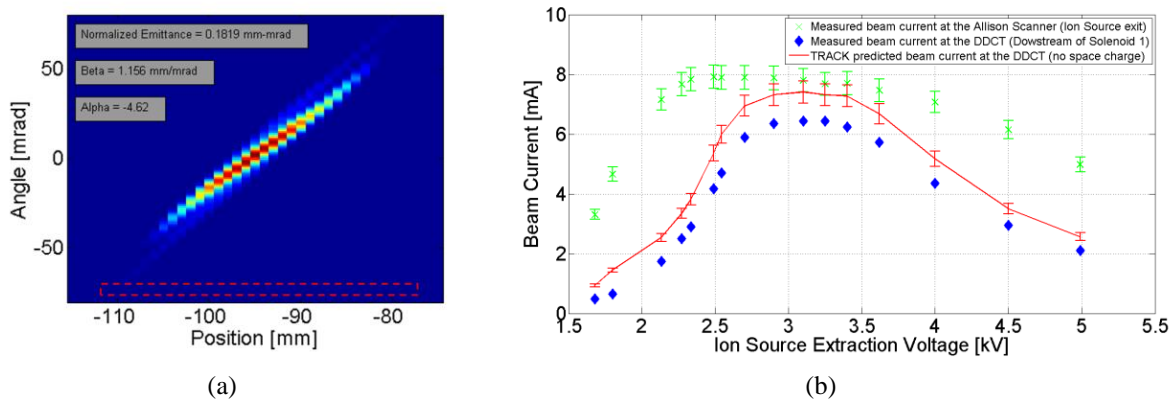

(a)             (b)

**FIGURE 5**. Examples of emittance and transmission measurements. (a) Phase portrait recorded near the ion source at the beam current (at the LEBT exit) of 10 mA. (b) Transmission between the ion source box and DCCT. The green crosses are the measured beam currents obtained from the phase space portrait integrals, and the blue diamonds are the beam currents measured by the DCCT and EID #1. The red curve is from TRACK simulations. The error bars reflect 5% error estimated for the image integral-to-current conversion coefficient. Ion source settings are optimized for 5 mA in the LEBT.

Also, Fig. 5b presents the sum of the currents measured by the DCCT and EID #1. It shows that at least ~20% of the beam exiting the ion source and as much as ~70% for low extraction voltages is being scraped off at the exit of the ion source box. No negative consequence of this scraping has been observed, since the ion source has enough overhead to provide up to 10 mA at the DCCT. As one can expect, TRACK simulations indicate that it noticeably decreases the beam emittance.

## RFQ matching

Emittance and Twiss parameters were measured with the emittance scanner located at the end of the LEBT, about 20 cm downstream of the virtual location of the RFQ vanes (see Fig. 2). Figure 6 reports the measured Twiss parameters as a function of the current in solenoid #3 for the beam current of 4.5 mA and 0.5 mA.

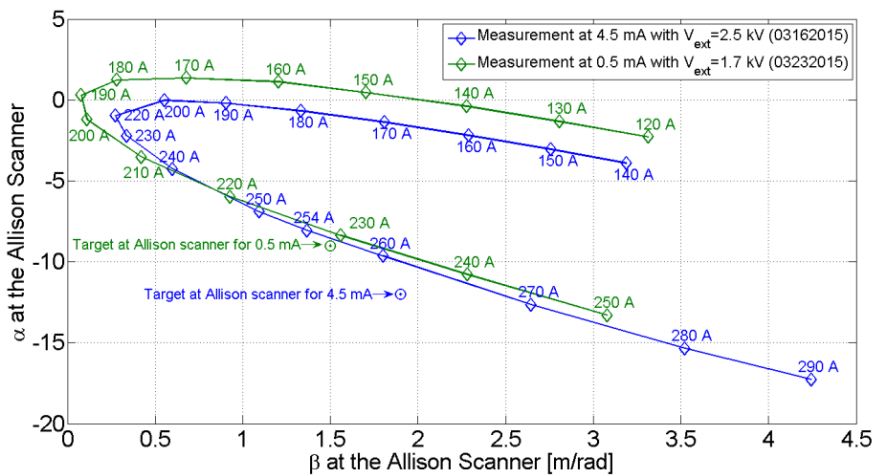

**FIGURE 6**. Relationship between the measured Twiss parameters at the end of the LEBT when varying the current of solenoid #3 for a 4.5 mA beam and 0.5 mA beam. Sampling is at the end of a 1 msec pulse chopped out of a 2 msec pulse from the ion source. Current in solenoids #1 and #2 are 141.4 A and 158 A respectively. The emittance measured at the target beta-function is 0.13 µm for the 4.5 mA and 0.11 µm for the 0.5 mA.

The design of the RFQ has been performed for a nominal input matched beam at the RFQ vanes of $\alpha$=1.6 and $\beta$=0.07 m and an input rms normalized emittance below 0.18 mm-mrad. After the 20-cm drift, these Twiss parameters translate to α=-12 and β=1.9 m for 4.5 mA and to α=-9 and β=1.5 m for 0.5 mA, defocusing from space charge accounting for the difference. In Fig.6, matching of the beta-function corresponds to deviation of alpha of ~10% at 0.5 mA and ~20 % at 4.5 mA.

Previous simulations discussed in Ref. [13] showed that the quality of the match into the PIP2IT RFQ tends to be more sensitive to the beam size (β) than the beam divergence (α). Recent TRACK simulations with a more realistic input distribution confirmed that a mismatch in alpha of about 20% does not dramatically impact the output transverse emittance of the RFQ.

After installation of the RFQ, matching is being done empirically, by measuring, first, the transmission efficiency, and then the beam quality (i.e. emittance) at the exit of the RFQ.

## Beam stability and stabilization loop

Throughout commissioning, we observed drifts of the beam Twiss parameters for seemingly identical ion source tunes and focusing settings. Concurrently, the beam current read out by the beam current monitors (DCCT, Faraday cup) may vary by as much as ~10% over long runs (Fig.7) or from day-to-day operation. The origin of either has not been understood, and no clear correlation was identified between the drifts of the current and Twiss parameters. In order to maintain constant beam conditions, a couple of stabilization loops were implemented.

First, the beam current can be kept constant by adjusting either the ion source extraction voltage. Second, Twiss parameters for matching to the RFQ can be kept nearly constant by monitoring the current lost to the LEBT scraper. If it deviates more than a predetermined fraction, the current of solenoid #3 is adjusted in small steps until the LEBT scraper lost current returns within its tolerance. Consequently, the Twiss parameters return close to their nominal value. Both loops were tested successfully, though whether it is worthwhile to implement them into routine operation of the PI2IT depends on details of the commissioning results of downstream sections.

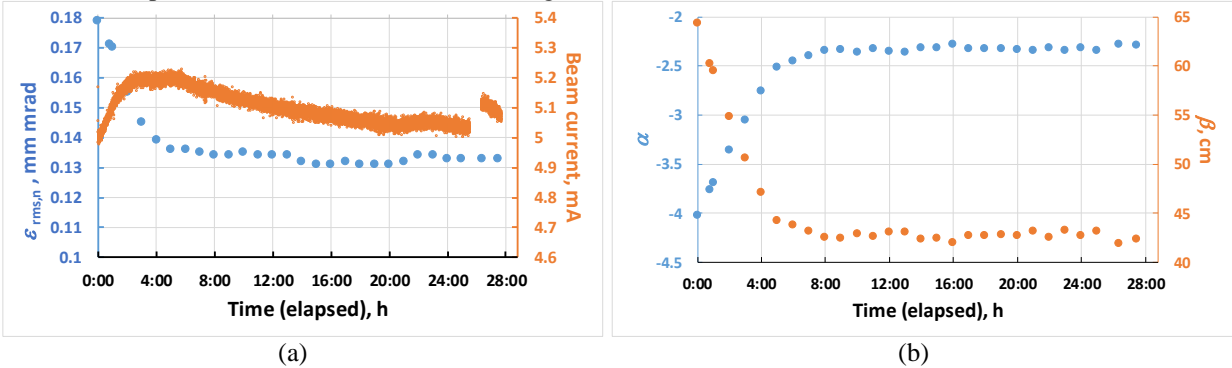

**FIGURE 7.** Drift of the beam parameters during a continuous run of a dc beam. (a) beam current and emittance, (b) Twiss parameters at the end of the LEBT. The discontinuity on the beam current trace is the result from a trip.

## Beam position and angle adjustment

The LEBT dipole correctors are located inside the solenoids. As a result, a change of a corrector current kicks the beam in both planes, with the amplitudes depending on the solenoid field. A 3D model of the LEBT solenoid together with the correctors was implemented in Microwave Studio in order to extract the 3D fields for implementation in TRACK. The input fields in the codes are normalized to match the measured solenoid and corrector field integrals. With solenoid fields typical for operation (3-5 kG), TRACK predicts the beam shift projected onto the horizontal or vertical axis to be 10 - 30% lower than the one for the same corrector outside of the solenoid.

In preparation for RFQ and MEBT commissioning, relationships between correctors currents that move the beam in pure horizontal or vertical plane were established with TRACK simulations for various solenoid currents. Figure 8a shows the agreement within 1% between the horizontal beam displacement measured with the emittance scanner and the TRACK predictions.

Similar TRACK simulations gave relationships between currents of the correctors in solenoids #2 and #3 that change only the horizontal or vertical angle in a given location without modifying the spatial position. Comparison with emittance scanner measurements (Fig. 8b) gives an accuracy for the angle prediction of ~10%.

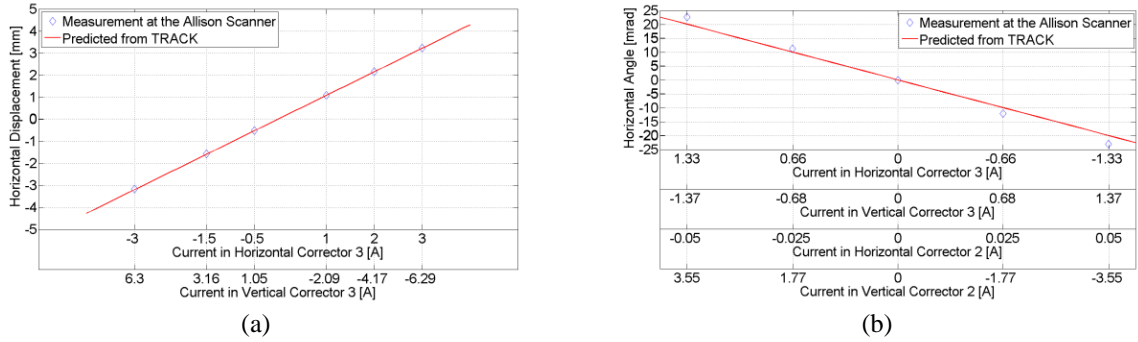

**Figure 8**. Testing of procedures for a horizontal shift (a) and an angle variation (b). Distance from the mid plane of solenoid #3 to the emittance scanner entrance is 415 mm. (a) Solenoid #3 current is 300 A. (b) Currents of solenoids #2 and #3 are 158 A and 253 A, correspondingly.

## MEASUREMENT OF SECONDARY ION PRODUCTION

Residual gas molecules can be ionized by the H⁻ beam ions and then be trapped in the negative beam potential, thus neutralizing the space-charge forces of the beam. The scheme chosen for the PIP2IT LEBT assumes nearly full neutralization upstream of Solenoid #2 and nearly no neutralization downstream (Fig. 1b), with biasing of EID#2 providing the separation between these two regions. To test these assumptions, the currents to the chopper plates were recorded for different conditions (Fig.9) with a 5 mA DC H⁻ beam. In these measurements, special attention was paid to ensure that there was no measurable losses of the primary beam in the vicinity of the chopper. The kicking plate was kept at -300 V, and the absorber was grounded through a measuring resistor. Currents to both plates were recorded as functions of the beam current and approximated by linear fits. Because secondary electrons knocked out of the kicking plate are accelerated toward the absorber, the ion current is calculated as $I_{ion} = I_{kicker} - |I_{absorber}|$.

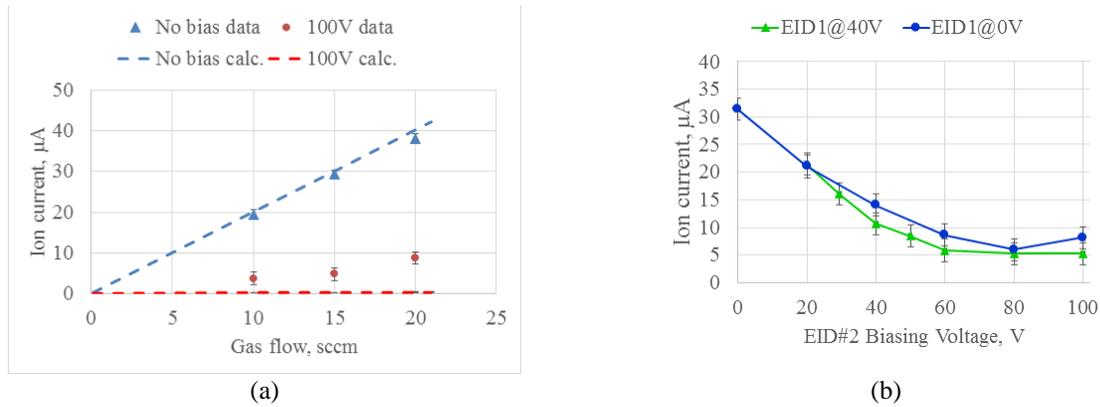

**FIGURE 9**. Measured current of positive ions as function of the hydrogen flow (a) and EID3 biasing voltage (b).

The LEBT vacuum is dominated by the hydrogen from the ion source, so that the pressure is proportional to the gas flow. The ion current is found to be proportional to the gas flow as well (Fig. 9a), supporting the assumption that the measured current is indeed created by positive secondary ions. For the case where EID#2 is not biased, the measured values fit well to calculations in the model of ions collected from the entire LEBT length between the ground electrode of the ion source and the LEBT scraper (that was kept biased at +50 V). For this calculation, we used the ionization cross section of $\sigma_i = 1.5 \times 10^{-16}$ cm² [14] and the gas density integral from vacuum simulations. With EID #2 biased to +100 V, the ion current drops drastically but still stays by an order of magnitude higher than the estimation for the flux created only between EID #2 and the LEBT scraper. A possible explanation of the discrepancy is a current load from secondary electrons created on the vacuum chamber walls near the chopper by fast neutrals and electrons from stripping. Taking this into account, the good level of agreement in the un-biased measurements is likely accidental.

The ion current is decreasing with the increase of the EID#2 potential (Fig.9b), coming to a steady state at ~60 V. With no neutralization inside EID#2, this voltage corresponds to nearly the ground potential at the beam axis, indicating that the space charge upstream is mostly neutralized. The ion current stays unchanged when biasing EID#1 from zero to +40 V. It can be interpreted as the beam potential inside the ground electrode still being higher than the one inside EID#1.

The results support the model where the secondary neutralizing ions are lost primarily longitudinally. This situation is possible when the typical longitudinal travel time, $\tau_{long} = L_{LEBT}/\upsilon_T \cong (1m)/(10^3 m/s) = 1 ms$ is comparable with neutralization time $\tau_n = (n_{gas}\sigma_i \upsilon_{H^-})^{-1} = 0.8\ ms$. Here $L_{LEBT}$ is a typical length of the neutralized area, $\upsilon_T$ is the thermal velocity of the positive ion, $n_{gas}$ is the residual gas density at $10^{-6}$ Torr, and $\upsilon_{H^-}$ is the velocity of the H$^-$ ions. In this scenario, the steady state potential along the beam is nearly constant and is determined by the potential at both ends of the neutralized section.

## FAST SECONDARY PARTICLES

Near the ion source where the hydrogen pressure is high, a measurable part of the H$^-$ ions is converted into either protons or neutral hydrogen atoms (H$^0$) by electron detachment in collisions with the residual gas molecules. These secondary particles have nearly the same speed and kinetic energy as the initial H$^-$ ions and can propagate through the entire LEBT. The detached electrons also have the same speed but their momentum is lower by the ratio of masses (~2000), so they are quickly separated from the main beam in the magnetic field of the dipole correctors and solenoids.

The neutrals fly ballistically, following the trajectories of primary H$^-$ ions. The protons are focused by solenoids just as H$^-$ ions are and, therefore, ignoring effect of imperfections and dipole correctors, fill in the same envelope as the H$^-$ beam. They can be accelerated in the RFQ and further, creating losses in bends or where the RF frequency changes [15].

One can estimate the portions of H$^-$ ions converted to neutrals and protons using the cross sections cited in [16] for 10 keV ($8\times10^{-16}$ cm$^2$ and $4.5\times10^{-17}$ cm$^2$, correspondingly). The integral of the hydrogen density over the length between the ion source and the first solenoid calculated from vacuum simulations is ~$1.4\times10^{18}$ m$^{-2}$ for the H$_2$ flow of 15 sccm. With these numbers, the portions of the initial H$^-$ beam converted into H$^0$ and protons are ~12% and ~0.6%, respectively.

Both protons and neutrals were observed experimentally. The proton beam is clearly seen in the emittance scanner. When the scanner was installed at the end of the LEBT, manipulations with dipole correctors allowed separating the H$^-$ and H$^+$ beams in phase space (as illustrated in Fig. 9 of Ref. [8]). In this mode, the relative intensity of the proton beam with respect to the H$^-$ was found to be between 0.2% and 0.6%. Similar result, 0.5%, was observed with the scanner installed near the ion source. Installation of the LEBT bend is expected to decrease this value by orders of magnitude.

Presence of neutrals is always obvious in the emittance scanner portraits as a signal independent of the voltage on the scanner plates. In normal scanner operation, with the suppressor electrode in front of the scanner collector set to $U_s$=-100V, the signal is small and is likely created by protons reflected from the surface of the scanner collector. The signal is greatly enhanced by secondary electron emission when $U_s$=0. With the scanner installed near the ion source, one can estimate the intensity of the neutrals from scanner portraits under two assumptions: (a) the secondary electron emission coefficient by the H$^-$ ions and the fast neutrals is the same, and (b) the phase distribution of neutrals in this location is identical to the one of the H$^-$ beam. Comparing the portraits recorded at $U_s$=0 and at its nominal value ($U_s$= -100V), the relationship between intensities of the H$^-$ images provides the secondary emission coefficient (found to be ~2). Then, in the portrait with $U_s$=0, the neutrals signal is summed over the position for a voltage on the plates (i.e. angle) far from the H$^-$ beam image (see the red dashed rectangle in Fig. 5a), and in the portrait with $U_s$=-100V, the beam signal is summed over the slit position for zero plate voltage. Comparison of the two sums, with the secondary emission taken into account, gives a relative intensity of neutrals of 8%.

An alternative measurement of the neutral flux is performed with calorimetry of an insulated electrode (an EID with no water-cooling) installed in one of the configurations, 48 cm downstream of the center of Solenoid #1. When a DC beam passes with no losses through the electrode's opening, the electrode's temperature increases linearly with time. Knowing the thermal capacity of the electrode (verified by direct deposition of the primary beam), one can deduce the power of the neutral beam portion intercepted at its surface. Assuming that the energy of all neutrals is 30 keV and the distribution of the neutrals is identical to the one of the primary beam with solenoid #1 turned off, the flux of neutrals coming to the EID is found to be 4% of the primary beam. The number agrees within 20% with both estimations and with the previously discussed measurements, if the neutrals divergence is taken into account.

Because of the LEBT 2-m length, the neutral flux at the RFQ entrance is negligible (estimated to be ~$8\times10^{-4}$ of the beam current) and will be completely eliminated with the installation of the LEBT bend.

## SUMMARY

The PIP2IT LEBT is successfully commissioned to its nominal parameters in a straight configuration, in both pulsed and DC modes of operation. For optimum tuning, the emittance measured at the LEBT exit at the nominal 5 mA beam current and Twiss parameters close to the ones specified for RFQ injection is 0.13 µm (rms, norm.). Pressure at the LEBT exit stays below $3\cdot10^{-7}$ Torr in all operational scenarios including with DC beam diverted onto the absorber. There is significant beam scraping in the upstream portion of the LEBT, ~20% for 5 mA beam current measured at the LEBT exit. It does not create operational problems and is beneficial since it decreases the emittance.

The current of positive ions, created by the beam from the residual gas molecules, to the chopper is found to be dependent on voltages applied to the electrode that separates the areas with high and low neutralization in the pursued scheme. The maximum collected current agrees with the estimation where the total positive ion production in the LEBT from the ground electrode to EID #2 is assumed.

Production of fast neutrals and protons due to electron detachment on the residual gas is estimated from the emittance scanner portraits and also, for neutrals, from calorimetric measurements. The experimental numbers agree well with calculations. In part, we estimate that ~0.5% of the beam is converted into the fast protons travelling through the entire LEBT in its straight configuration.

## ACKNOWLEDGMENTS

Authors acknowledge the initial work at LBNL on the LEBT design made by J. Staples and characterization of the ion source by Q. Ji; participation in the LEBT commissioning by J. Steimel and F. Garcia; and writing a MathCad program for quick LEBT optics simulations by V. Lebedev. We are thankful to a large team that designed and assembled various subsystems, including G. Saewert (chopper and modulator), R. Brooker, K. Carlson (electrical systems), M. Kucera (controls), A. Saewert, A. Ibrahim (diagnostics), R. Andrews, C. Baffes, T. Hamerla, K. Kendziora, D. Lambert, D. Snee (mechanical).